\documentclass[useAMS,usenatbib]{mn2e}

\usepackage{graphicx}
\usepackage{amsmath}
\usepackage{amssymb}

\newcommand{\somme}[2]{\underset{#1}{\overset{#2}{\sum}}}
\newcommand{\eg}[0]{\;=\;}
\newcommand{\ega}[0]{& = &}
\newcommand{\x}[0]{\times}
\newcommand{\abs}[1]{\mid{#1}\mid }
\newcommand{\integ}[2]{\int\limits_{#1}^{#2}}
\newcommand{\pl}[1]{P_{#1}}

\newcommand{\ylm}[2]{Y_{#1}^{#2}}
\newcommand{\alm}[2]{a_{{#1}{#2}}}
\newcommand{\cl}[1]{C_{#1}}
\newcommand{\clest}[1]{\widehat{C}_{#1}}
\newcommand{\pcl}[1]{\widehat{C}_{#1}^p}
\newcommand{\tcl}[1]{\widehat{C}_{#1}^t}
\newcommand{\fcor}[0]{\xi}
\newcommand{\fcorp}[0]{\widetilde{\xi}}
\newcommand{\fcorest}[0]{\hat{\xi}}
\newcommand{\pfcor}[0]{\hat{\xi}^p}
\newcommand{\tfcor}[0]{\hat{\xi}^t}
\newcommand{\teafi}[0]{(\theta,\phi)}
\newcommand{\CVec}[1]{ \left[ #1 \right] }
\newcommand{\CMtx}[1]{ \left[ \left[ \mathbf{#1} \right] \right] }

\usepackage{color}

\newcommand{\modif} {  } 

\newcommand{\grfile}[1] {#1.eps} 

\title[Partial CMB maps: bias removal, optimal binning of $\cl{\ell}$ ]
{Partial CMB maps: bias removal and optimal binning of the angular power spectrum}

\author[R. Ansari and C. Magneville]{R. Ansari$^{1,2}$ \thanks{E-mail:
ansari@lal.in2p3.fr (RA); christophe.magneville@cea.fr (CM)} and C. Magneville$^{3}$ \\ 
$^{1}$Universit\'e Paris-Sud, LAL, UMR 8607, F-91898 Orsay Cedex, France\\
$^{2}$CNRS/IN2P3,  F-91405 Orsay, France\\
$^{3}$CEA, DSM/IRFU, Centre d'Etudes de Saclay, F-91191 Gif-sur-Yvette, France}
\begin{document}

\date{Accepted XXXX , Received 2009 October 10; in original form 2009 October 10}

\pagerange{\pageref{firstpage}--\pageref{lastpage}} \pubyear{2010}

\maketitle

\label{firstpage}

%%%%%%%%%%%%%%%%%%%%%%%%%%%%%%%%%%%%%%%%%%%%%%%%%%%%%%%
%%%%%%%%%%%%%%%%%%%%%%%%%%%%%%%%%%%%%%%%%%%%%%%%%%%%%%%

\begin{abstract}
We present a semi-analytical method to investigate the systematic effects and statistical 
uncertainties of the calculated angular power spectrum when incomplete spherical maps 
are used. The computed power spectrum suffers in particular a loss of angular frequency 
resolution, which  can be written as $\delta \ell \sim \pi/\gamma_{max}$, where $\gamma_{max}$ is 
the effective maximum extent of the partial spherical maps. We propose 
a correction algorithm to reduce systematic effects on the estimated $C_\ell$, as 
obtained from the partial map projection on the spherical harmonic $\ylm{\ell}{m}$ basis.  
We have derived near optimal bands and weighting functions in $\ell$-space for power 
spectrum calculation using small maps, and a correction algorithm for partially masked 
spherical maps that contain information on the angular correlations on all scales.
\end{abstract}

\begin{keywords}
methods: data analysis -methods:statistical -techniques:image processing 
cosmology: cosmic microwave background
\end{keywords}

%%%%%%%%%%%%%%%%%%%%%%%%%%%%%%%%%%%%%%%%%%%%%%%%%%%%%%%
%%%%%%%%%%%%          Section I - Introduction            %%%%%%%%%%%%%
%%%%%%%%%%%%%%%%%%%%%%%%%%%%%%%%%%%%%%%%%%%%%%%%%%%%%%%

\section{Introduction}
The measurement of the temperature and polarisation anisotropies of the
Cosmic Microwave Background (CMB) radiation provides essential information for testing
the cosmological models and determining their parameters. Most of the statistical information 
present in the CMB temperature or polarisation sky maps can be encoded in the angular power 
spectrum $\cl{\ell}$. An overview of the physical mechanisms responsible for the CMB anisotropies 
and the effect of the cosmological parameters on the power spectrum shape can be found in 
\citep{cmbcosmop} and \citep{introcmb}. 

The following references give an overview of some recent CMB power spectrum measurements:
The all sky WMAP (Wilkinson Microwave Anisotropy Probe) space mission (\cite{wmap5yps}, \cite{wmap7yps});
ACBAR  \citep{acbar} which has made high resolution measurements using the Viper 
telescope at South Pole Station; the South Pole Telescope - SPT \citep{spt};
the Cosmic Background Imager-CBI \citep{cbi}
and the ACT telescope \citep{act} in the Atacama desert;
the polarisation measurement at Amundsen-Scott South Pole Station by the 
Degree Angular Scale Interferometer - DASI \citep{dasi}; as well as ballon borne experiments
such as BOOMERANG (\cite{boomerang1} , \cite{boomerang2}), 
Archeops (\cite{archeops1}, \cite{archeops2}) and MAXIMA \citep{maxima}.
Table 1 summarizes typical sky coverage, the accessible $\ell$ range, and microwave 
frequency range for some of the above  mentioned instruments.
\begin{table*}
 \centering
  \caption{$\ell$-range and sky coverage for few of the microwave sky observation instruments}
  \begin{tabular}{|l|cccc|}
  \hline
   Name     & sky cov. (deg$^2$)  &  Angular Resolution (arcmin) & $\ell$ range (min-max) & frequency coverage (GHz) \\
 \hline
   WMAP & $\sim$ 40 000 (4 $\pi$ sr)  & 10-60 & 1-1500 &  23-94 \\
   ACBAR & 600 (10 fields) & 5 & 200-3000 & 150,220 \\
   BOOMERANG & 750 & 6-10 & 10-1500 & 145,245,345 \\
   Archeops & 5000 & 11 (@143) & 10-700 & 143,217,353,545 \\
   CBI & 40 (3 fields) & 5-10 & 200-3500 & 26-36 \\
   SPT & 100 & $\sim$ 1 & 2000-9500 & 150,220 \\
   ACT & 228 & 1.4 & 600-8000 & 148 \\  
 \hline
\end{tabular}
\end{table*}

Determining the CMB angular power spectrum 
is a complex process  and the data analysis must take into account a number of effects, such as 
the non-stationnary noise contribution, non-circular instrumental beams, the sky scanning 
strategy and foreground contamination. A review of the general methods for CMB data processing 
can be found, for example, in \citep{tristram}.

The different systematic and statistical effects on CMB angular power specrtum estimates
from partial sky maps have already been studied in length.  In particular, several 
Montecarlo or analytical methods have been devised  to evaluate or correct the impact of
limited sky coverage on estimated angular power spectrum (\cite{master},\cite{mitra},\cite{dasperg}).

In this paper we propose a different approach which is based on the analysis
of the distortion of the angular correlation function. However, it should be noted that 
explicit computation of the angular correlation function is not needed.
The use of the correlation function has also been studied by several authors, 
although with a different approach than the one developed here (\cite{szapud1} , \cite{szapud2}). 
In section 2, we recall briefly the main relations between the angular correlation 
function $\fcor(\gamma)$ and the angular power spectrum $\cl{\ell}$, and express them 
as a set of linear algebraic equations. Using this formalism in section 3, 
we compute the distortion of the estimated $\cl{\ell}$ in incomplete sky maps 
and we show how the angular correlation function can be corrected to minimize 
these distortions. We show in particular that power spectrum estimates on partial 
maps suffer a loss of resolution in $\ell$-space and we propose near optimal 
$\ell$-space window function (binning).

The algebraic approach presented in section 3  can not be used to compute the variance 
of the reconstructed angular power spectrum (see paragraph 3.1 and 3.4). 
We have thus used Montecarlo simulations to estimate $\cl{\ell}$ uncertainties. 
We present in section 4 the corresponding results for 
small maps   ($\Omega \simeq 4 \pi \times 10^{-2}$), representative of ground or balloon experiments
such as OLIMPO \citep{olimpo}, as well as the systematic shifts and possible 
correction for nearly complete  maps ($\Omega \simeq 4 \pi \times 0.9$) representative of space missions such as Planck
\citep{Planck}.

%%%%%%%%%%%%%%%%%%%%%%%%%%%%%%%%%%%%%%%%%%%%%%%%%%%%%%%
%%%%%%%%%%%%          Section II - C(l) <> FCor           %%%%%%%%%%%%%
%%%%%%%%%%%%%%%%%%%%%%%%%%%%%%%%%%%%%%%%%%%%%%%%%%%%%%%

\section[]{Angular correlation function and power spectrum from the full sphere}

We recall here the basic relations for the angular power spectrum
and correlation function. Detailed derivation of most of theses formulae 
can be found in \citep{cmvcl} or \citep{wand1}.

We consider a real signal $s(\vec{\Omega})$
on the sphere ($S^2$), measured for each direction $\vec{\Omega} = (\theta, \phi)$. 
The function $s$ can be expanded on the spherical harmonic basis:
% \begin{eqnarray*}
%  s(\theta, \phi) & = & \sum_{l=0}^{\infty} \sum_{m=-\ell}^{m=+\ell} 
% a_{\ell m} \, Y_\ell^m (\theta, \phi) \\
% a_{\ell m } & = & \int s(\vec{\Omega}) Y_\ell^m (\vec{\Omega}) d \Omega \\
%\end{eqnarray*}
\begin{eqnarray}
 s \teafi & = & \somme{l=0}{\infty} \somme{m=-\ell}{m=+\ell} 
                \alm{\ell}{m} \, \ylm{\ell}{m} \teafi  \label{s2alm} \\
\alm{\ell}{m} & = &
              \integ{S^2}{} s(\vec{\Omega}) \, (\ylm{\ell}{m} (\vec{\Omega}))^{*} d \Omega  \label{alm2s} 
\end{eqnarray}

For an isotropic random signal, the angular power spectrum $\cl{\ell}$ characterizes 
the statistical properties of the signal ($ \langle \rangle $ denotes ensemble average). 
\begin{eqnarray}
\langle \alm{\ell}{m}  \alm{\ell'}{m'}^* \rangle & = & \delta_{\ell-\ell'} \delta_{m-m'}  \cl{\ell}  
\end{eqnarray}
It is possible to compute an unbiased power spectrum estimator
$\clest{\ell} $ from the spherical harmonic expansion coefficients. The 
$\clest{\ell} $ coefficients are independent random variables with variance 
$\sigma^2_{\clest{\ell} } $ (cosmic variance):
\begin{eqnarray}
\clest{\ell} & = & \frac{1}{2 \ell + 1} \somme{m=-\ell}{m=+\ell} \abs{\alm{\ell}{m} }^2 \label{clfralm} \\
\langle \clest{\ell}  \rangle & = & \cl{\ell}   \notag \\
\sigma^2_{\clest{\ell} }  & = & \frac{2}{2 \ell + 1} \cl{\ell}^2  
\end{eqnarray}
For an isotropic signal, the  angular correlation function $\xi(\gamma)$ can also be used to characterize 
the signal properties, where $\gamma$ is the separation angle ($ \cos \gamma = \vec{\Omega} \cdot \vec{\Omega'} $) : 
\begin{eqnarray}
\fcor(\gamma) & = & \frac{1}{4 \pi} \somme{l=0}{\infty} (2 \ell + 1 ) \cl{\ell} \pl{\ell}(\cos \gamma) 
\label{ksifrcl} \\
\cl{l} & = & 2\pi\;\integ{-1}{+1}\;\fcor(\gamma)\;\pl{l}(\cos \gamma )\;d\cos \gamma \label{clfrksi}
\end{eqnarray}

It should  be noted that the above relation holds also for the computed angular 
correlation function ($\fcorest(\gamma)$) and power spectrum ($\clest{\ell}$) 
for a given sky realization.
\begin{eqnarray}
\fcorest(\gamma) & = & \frac{1}{4 \pi} \somme{l=0}{\infty} (2 \ell + 1 ) \clest{\ell} \pl{\ell}(\cos \gamma) \label{ksiestfrcl} \\
   & = & \frac{1}{ \mathcal{N} (\gamma)  } \integ{S^2 \times S^2}{} s(\vec{\Omega}) \,  s(\vec{\Omega'}) \notag 
\delta \left(  \cos \gamma - \vec{\Omega} . \vec{\Omega'} \right) d \Omega d \Omega' \\
\mathcal{N} (\gamma)  & = & \integ{S^2 \times S^2}{} 
\delta \left(  \cos \gamma - \vec{\Omega} . \vec{\Omega'} \right) d \Omega d \Omega' = 8 \pi^2  \notag
\end{eqnarray}
The above relations, analogous to Fourier series, contain algebraic (sum) and analytic
(integral) expressions, and involve  the discrete variable  $ \ell \in \mathbb{N} $,
as well as the continuous variable $\gamma \in [0, \pi] $. We can rewrite these expressions in a purely
algebraic form, similar to the Discrete Fourier Transform (DFT).  
Indeed, if the $\cl{l}$ spectrum is negligible for large $\ell > \ell_{max}\;$, 
and for any set of discrete values $\{ \gamma_i \}$ of the $\gamma$ angle, 
the relations~\ref{ksifrcl} and~\ref{ksiestfrcl} can be written in matrix form:
\begin{equation}
\CVec{ \fcor(\gamma_i) } \eg \CVec{ \fcor_i } = \CMtx{K_{i \ell} } * \CVec{\cl{\ell}} \label{ksifrclmat} 
\end{equation}
% due for example to the intrumental lobe
In the particular case where the number of $\gamma_i$ values is equal to the $\ell_{max}+1$ non-zero
$\cl{\ell}$ values, the $\CMtx{K_{i \ell} }$ matrix would be square. It can then be 
shown that the separation angles $\gamma_i$ at which the correlation function $\fcor$ is computed 
($\fcor_i=\fcor(\gamma_i)$), could be choosen such that the square 
matrix $\CMtx{K}$ of size $(\ell_{max}+1) \times (\ell_{max}+1)$ is non-singular
(see appendix). % ~\ref{appmatinv}). 
The equation~(\ref{ksifrclmat}) may thus be inverted to get:
\begin{equation}
\CVec{\cl{l}} \eg \CMtx{K}^{-1} * \CVec{\fcor_i}  \label{clfrksimat}
\end{equation}
Moreover, these $\gamma_i$ values are not very different from 
\mbox{$\ell_{max}+1$} equidistant $\gamma$, distributed from $0$ to $\pi$, with the matrix 
elements close to  $\CMtx{K}^{-1}_{\ell,i} \propto \pl{\ell}(\cos \gamma_i)$.

%%%%%%%%%%%%%%%%%%%%%%%%%%%%%%%%%%%%%%%%%%%%%%%%%%%%%%%
%%%%%%%%%%%%                     Section III                    %%%%%%%%%%%%%
%%%%%%%%%%%%%%%%%%%%%%%%%%%%%%%%%%%%%%%%%%%%%%%%%%%%%%%

\section[]{Angular power spectrum from partial maps}

%%%%%%%%%%%%%%%%%%%%%%%%%%%%%%%%%%%%%%%%%%%%%%%%%%%%%%%
\subsection[]{Angular correlation function distortion}
The above equation~(\ref{clfrksimat}) can be used to analyse the impact of
any linear distortion of $\fcor$ on the calculated $\cl{l}$. 
Any linear distortion of the angular correlation function (including truncation for 
$\gamma > \gamma_{max}$) can be represented by a matrix $\mathbf{[[D]]}$ applied to $\xi$. 
\begin{eqnarray}
\langle \CVec{ \fcor(\gamma) }^d \rangle \ega \CMtx{D} * \CVec{ \fcor(\gamma) } \notag \\
\langle \CVec{\cl{l}}^d \rangle \ega \CMtx{K}^{-1} *  \CMtx{D}  * \CMtx{K}  * \CVec{\cl{l}} \notag \\
\langle \CVec{\cl{l}}^d \rangle \ega \CMtx{B_{c}} * \CVec{\cl{l}} \notag \\
\CMtx{B_{c}} \ega \CMtx{K}^{-1} *  \CMtx{D}  * \CMtx{K} \label{bctdef}
\end{eqnarray}
It should be stressed that in most cases, we can only compute the mean  distortion 
caused by the measurement process, in particular due to the incomplete coverage.
The above relations will not hold in general for a given realization 
or a single measurement. It will only be valid when an ensemble average 
is taken for the angular correlation function and power spectrum. This explains 
why it can not be used to estimate statistical uncertainties on the computed 
power spectrum.

Using the formalism described above, we have computed the effect of distorting
or modifying the angular correlation function in some typical cases such as:
\begin{itemize}
\item Application of a sharp or a smooth cut $\xi(\gamma) \rightarrow \xi ^d(\gamma) = \xi(\gamma) * \mathrm{Cut}(\gamma)$
to the angular correlation function. In the case of a sharp cut, 
$\xi(\gamma)$  is set to zero for $\gamma > \gamma_{max}$ (step function cut).
\item Undersampling $\xi_i$ by a factor $p > 1$, {\em i.e.} $\xi_i$ known only for $i = k \times p$ 
\item Binning effect, where $\xi(\gamma_i)$ is replaced by the mean value 
of $\xi$ in a small interval around  $\gamma_i$. The binning can represent 
the distortion of the $\xi$ when computed through the histogram
of all pixel pair products $s(\vec{\Omega}) s (\vec{\Omega'})$, binned
as a function of their separation angle $\gamma$ ($\vec{\Omega} \cdot \vec{\Omega'} = \cos(\gamma) $).
\end{itemize}

\begin{figure*}
\begin{minipage}{\textwidth }
\includegraphics[width=\textwidth]{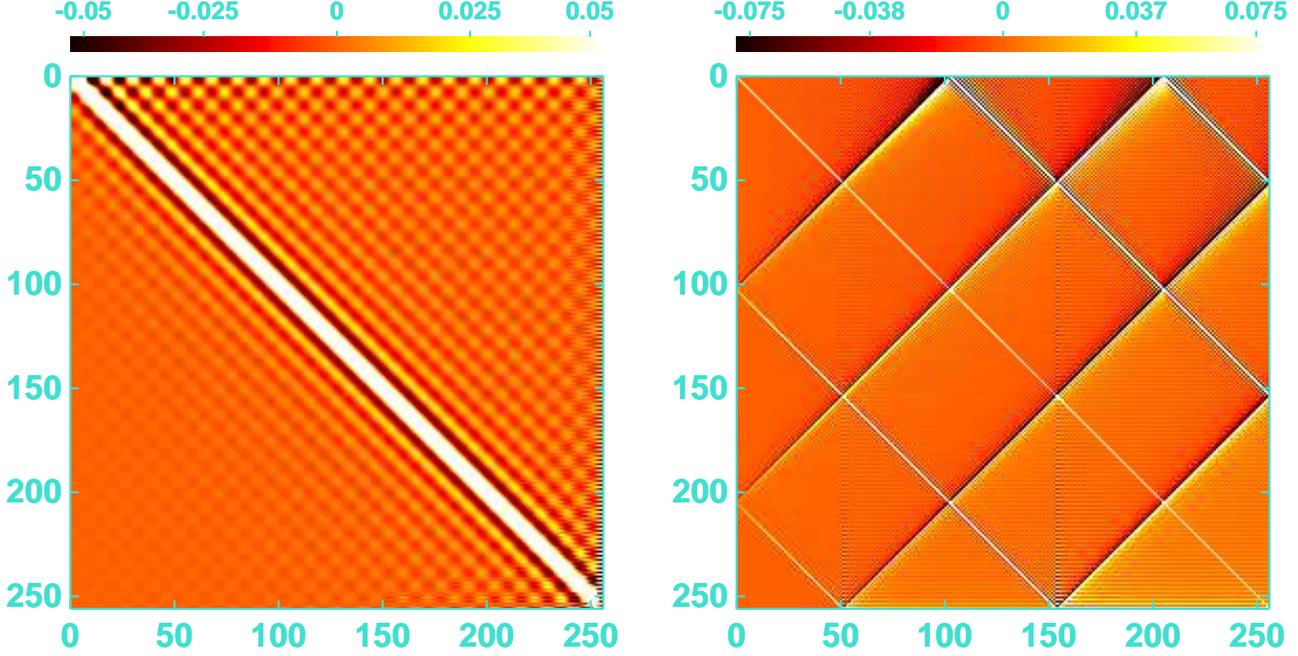} 
 \caption{Color scale representation of the $\CMtx{B_{c}}$ matrix relating $\langle \clest{\ell}^t \rangle \leftrightarrow \cl{\ell}$.
Left:  Correlation effect induced by restricting the angular range to $\gamma < 30^\circ$. Right: aliasing effect when 
$\xi$ is undersampled by factor p=5.  The color scale has been chosen to enhance visually the matrix structure (the 
diagonal terms are around $\sim 0.15-0.2$). }
 \label{kkinvbct}
\end{minipage}
\end{figure*}

Figure~\ref{kkinvbct} illustrates the effect of restricting the $\gamma$ range to $\gamma_{max}=30^\circ$
({\em i.e.} setting $\xi(\gamma)=0$ for $\gamma > \gamma_{max}$), (left) and the undersampling (right).
Restricting the range of angles for $\xi(\gamma)$,  creates a correlation between different $\ell$ while 
undersampling produces an aliasing effect.
Both of these effects are analogous to well known effects in standard Fourier analysis. 

%%%%%%%%%%%%%%%%%%%%%%%%%%%%%%%%%%%%%%%%%%%%%%%%%%%%%%%
\subsection{Partial maps : truncated $\xi^t(\gamma)$ } 
When the angular correlation function is calculated from maps with a maximal extent
$\gamma_{max}$, nothing can be known on the correlation function for $\gamma > \gamma_{max}$.
In addition, the statistical errors on the estimated correlation function will be larger 
compared to the one computed on the corresponding full ($4 \pi$) map. 
It is well known that computing the angular power spectrum from a truncated $\xi^t(\gamma)$ 
using the integral equation~(\ref{clfrksi}) or the linear combination equation~(\ref{clfrksimat})
produces spurious oscillations. However, these 
oscillations can be filtered out if the resulting power spectrum is binned.   
\begin{eqnarray}
\langle \clest{\ell}^t \rangle \ega \CMtx{B_{c}} * \CVec{ \cl{\ell} } \\
\clest{L}^t \ega \sum_\ell w_L(\ell)    \clest{\ell}^t  
\end{eqnarray}
The filtered or weighted power spectrum $\clest{L}^t $ defined here is obtained by applying the weight 
function $ w_L(\ell) $ to the power spectrum $\clest{\ell}^t $. The weight function should be centered around 
$L$ and normalised such that  $\somme{l}{}\;w_L(\ell)\;=\;1$. The $w_L(\ell)$ would be in general positive, 
with a maximum for $\ell = L$  and decreasing to zero ( $w_L(l)\rightarrow 0$) when $\abs{\ell-L}$ increases.
The expectation value of the filtered spectrum can then be expressed using the weight matrix 
$ \CMtx{W}_{L, \ell} =w_L(\ell) $: 
\begin{eqnarray}
\langle \clest{L}^t \rangle \ega \CMtx{W} * \CMtx{B_{c}} * \CVec{ \cl{\ell} }  \\
\langle \clest{L}^t \rangle \ega \CMtx{B_{cW}} * \CVec{ \cl{\ell} }  
\end{eqnarray}

\begin{figure}
\includegraphics[width=0.5\textwidth]{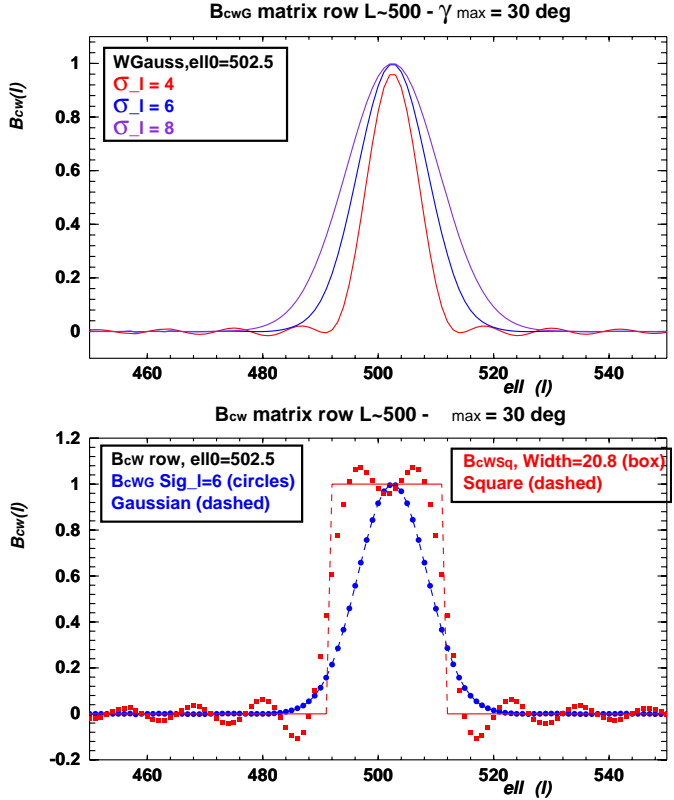} 
 \caption{ Weighted $\clest{L}^t$ calculation truncated $\xi$ at $\gamma_{max}=30^\circ$.
Top: Row of the weighted Bct matrix ($\CMtx{B_{cW}}$) 
around $L \sim 500$ for Gaussian weights ($B_{cWG}$) with three 
different widths $\sigma_\ell = 4,6,8$. Bottom: Comparison of 
$B_{cW}$ matrix row for $L \sim 500$ 
with the corresponding Gaussian weights ($\sigma_\ell = 6$), 
in blue and top hat (square) weighted ($B_{cWSq},  \mathrm{width}=6\times\sqrt{12}=20.8$) in red.  }
\label{rowbctw1}
\end{figure}

We have computed the matrix $\CMtx{B_{cW}}$ relating
true expectation values of $\cl{\ell}$ to mean value of the weighted power spectrum
$\langle \clest{L}^t \rangle $ from truncated angular auto-correlation function. The results 
shown here correspond to $\fcor(\gamma)$  truncated above $\gamma_{max}= 30^\circ$, for Gaussian or 
square (step-wise) weight  functions. Figure~\ref{rowbctw1} shows one of the rows of the $\CMtx{B_{cW}} $
matrix around $\ell \sim 500$ for different weight functions. 
For Gaussian weights width $\sigma_\ell \gtrsim \dfrac{\pi}{\gamma_{max}}$, the effective 
$\ell$-space window or filter function becomes numerically very close 
to the corresponding Gaussian function. It can be seen also that 
applying top hat or square weights result in an effective window function 
significantly different from the corresponding square function. 
We have checked that this property and the value of optimal 
weight function width $\sigma_\ell\simeq \dfrac{\pi}{\gamma_{max}}$ does not 
depend on the value of the central $L$, at least for 
large enough $L \gtrsim \sigma_\ell$. 

%%% Figure supprime -> texte enleve  (29/01/2010 ) 
% Figure~\ref{diffbctwgsq} shows deviation $B_{cWG} - W_{G} , B_{cWSq}-W_{Sq}$ 
% of the obtained  window function in the true $\ell$-space, from the corresponding weight
% functions  (Gaussian, square) when weighted power spectrum is computed the 
% truncated angular correlation function ($\xi=0$ for  $\gamma > 30^\circ$), 
% for two Gaussian weights ($\sigma_\ell=4,6$) and square weight. 
% It can be seen that for $\sigma_\ell =6$, the difference between 
% the effective window function and the corresponding Gaussian
% function is less than $5 \,\x\, 10^{-3}$. 

Our computation and simulations suggest that a Gaussian function 
with a width $\Delta \ell \sim \sigma_{\ell} = \dfrac{\pi}{\gamma_{max}}$ is a near optimal 
binning when the angular power spectrum is 
calculated from partial maps with maximum 
angular extent $\sim \gamma_{max}$. 
{\modif 
The word ``optimal'' should be understood in its common sense, and not the mathematical one, 
as the optimal solution depends on the chosen quantitative criteria. 
For example, different binning should be used if one seeks to increase the spectral resolution
or if one is concerned by the statistical errors. As explained above, the suggested Gaussian 
binning  has the following properties: 
\begin{itemize}
\item The Gaussian weighted corrected angular spectra reduces the window function 
tails by an order of magnitude, compared to pseudo-$\cl{\ell}$, while  maintaining
the $\ell$-space resolution as well as similar statistical  uncertainties 
(see section \ref{Montecarlo}). Using top hat (square) binning yields a window function 
 with long tails and  oscillations. 

\item The resulting true $\ell$-space window function is nearly identical to the weight or filter function 
applied to the computed $\cl{\ell}$, with a simple analytical expression (Gaussian). 
This property is useful for presenting measured power spectra. 
\end{itemize}
}

%%% Figures supprimee du papier
% \begin{figure}
% \includegraphics[width=0.5\textwidth]{\grfile{diffbctwgsq}} 
% \caption{Difference between the true $\ell$-space window function, 
% from the corresponding the Gaussian / square weight function ($B_{cWG}-W_{G}, B_{cWSq}-W_{Sq}$) 
% for truncated $\xi$ at $\gamma_{max}=30^\circ$.  Top: Gaussian weights, $\sigma=\pi/\gamma_{max}=6$, 
% middle $\sigma=4$, bottom square weights width= $6\times\sqrt{12}=20.8$. }
% \label{diffbctwgsq}
% \end{figure}

% \begin{figure}
% \includegraphics[width=0.5\textwidth]{\grfile{bwgsq}} 
% \caption{Matrix $BWG$ (top) and $BWSq$ relating binned $\tilde{C}_L$ 
% estimates to true $C_\ell$ ,  $\tilde{C}_L = BWG * C_\ell , BWSq * C_\ell$,
% for Gaussian and square type weights }
% \label{bwgsq}
% \end{figure}

When the power spectrum is estimated  using incomplete spherical maps,
a loss of resolution in angular frequencies ($\Delta\ell\sim \pi/\gamma_{max}$)
can also be understood using 
linear combinations of Legendre polynomial. We define new functions $Q_L$,
through linear combinations of Legendre polynomials around a given $L$, with Gaussian 
weights :
\begin{equation}
\begin{array}{lcl}
Q_{L}^{\sigma_l} & = & \frac{1}{\Sigma w} \sum_l  
  \exp \left[ \frac{(\ell - L)^2}{2 \sigma_\ell^2} \right] \, P_\ell \\
\Sigma w & = & \sum_l \exp\left[ \frac{(\ell - L)^2}{2 \sigma_\ell^2} \right] \\
Q_{L}^{\sigma_l} (\cos \gamma) & \simeq & P_{\ell} (\cos \gamma) 
\hspace{5mm} \gamma < \pi/\sigma_\ell \\
Q_{L}^{\sigma_\ell } (\cos \gamma) & \simeq & 0  \hspace{5mm} \gamma > \pi/\sigma_l
\end{array} 
\end{equation}

\begin{figure}
\includegraphics[width=0.5\textwidth]{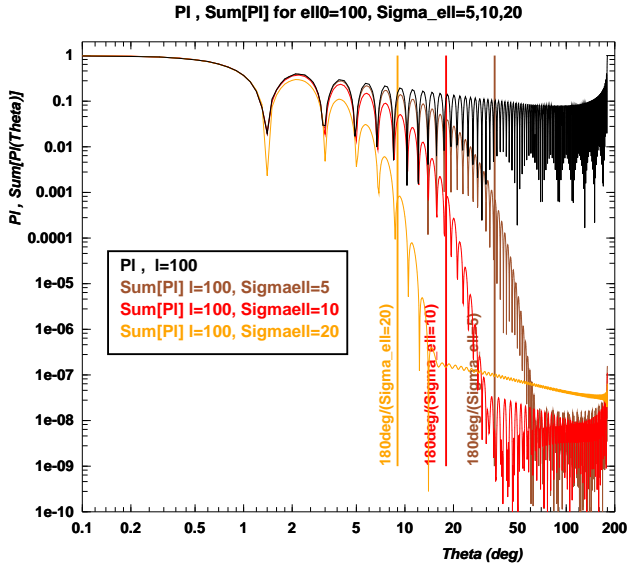} 
 \caption{Legendre Polynomial and $Q_L = \sum P_\ell$ functions, plotted for $L = 100$
 and $\sigma_\ell = 5,10,20$. We have plotted $| P_\ell |, | Q_L | )$.
 Notice also that both axes have logarithmic scales. }
  \label{figsompl2}
\end{figure}

\begin{figure}
\includegraphics[width=0.5\textwidth]{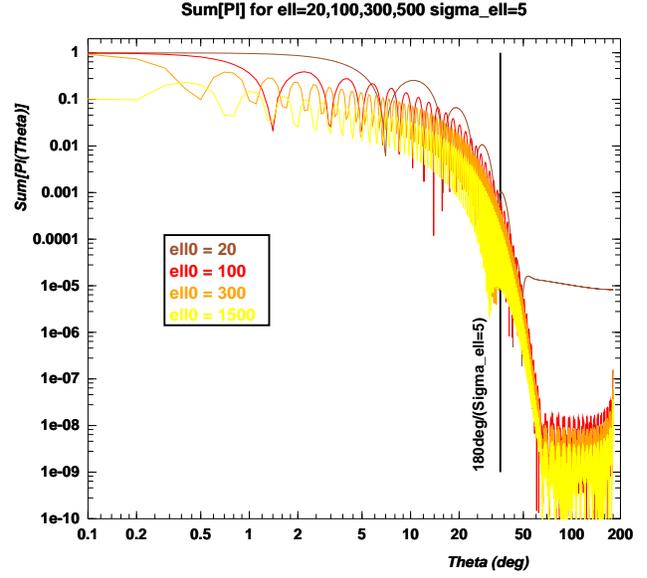} 
 \caption{Legendre Polynomial $| P_\ell |$ and and the $| Q_L |$functions, plotted for 
 $\sigma_\ell = 5$, and $\ell 0 = 20,100,200,1500$.
 Notice  that both axes have logarithmic scales. }
\label{figsompl1}
\end{figure}

Note that these new functions $Q_L(\cos \gamma)$ behave like 
$P_L(\cos \gamma)$ at small angles ($\gamma \lesssim \gamma_{max}=\pi/\sigma_\ell$) and become 
negligible at large angles ($\gamma \gtrsim \gamma_{max}=\pi/\sigma_\ell$). This property is 
illustrated on the figures~\ref{figsompl2}  and~\ref{figsompl1}. 
It might be possible to use these $Q_L$ functions as a basis
to decompose truncated angular correlation functions.

% \begin{figure}
% \includegraphics[width=0.5\textwidth]{\grfile{sompl3}} 
%  \caption{Legendre Polynomial and $Q_\ell$ computed up to $\pm n_s \times \sigma_l$ 
% with $n_s=1,3,5$, as well as a linear combination of $P_l$ with a square weight function. }
% \label{figsompl3}
% \end{figure}

%%%%%%%%%%%%%%%%%%%%%%%%%%%%%%%%%%%%%%%%%%%%%%%%%%%%%%%
\subsection[]{Partial maps : Decomposition on the $Y_\ell^m$ basis }
As we mentioned in section 2, a signal defined as a function 
of direction can be decomposed in spherical harmonics
(see equation~(\ref{s2alm})). Optimized algorithms for performing 
numerically this decomposition on pixelized spherical maps 
are commonly used in analyzing CMB data (\cite{clfast} , \cite{healpix}). 
The computed $\alm{\ell}{m}$ coefficients are used to derive the angular power spectrum $\clest{\ell}$ 
using equation~(\ref{clfralm}). 
When an incomplete (or partial) sky map is expanded on the $Y_\ell^m$ 
basis, the resulting power spectrum suffers systematic effects and is
often called a $\mathrm{pseudo-}\cl{\ell}$ spectrum.  The $\mathrm{pseudo-}\cl{\ell}$ spectrum distortion has 
already been studied by several authors and can be found for example in (\cite{master}, \cite{cmvcl}).
The partial map can be written as the result of applying a mask on the original signal. 
One can then show that the $\tilde{a}_{\ell m}$ coefficients computed on the masked
map can be related to the true  $\alm{\ell}{m}$ through a linear relation, which can be 
written in matrix form:
\begin{eqnarray}
s^p \teafi \ega s \teafi \times \mathrm{mask} \teafi  \\
\CVec{ \tilde{a}_{\ell m}   } \ega \CMtx{A} * \CVec{ a_{\ell m}  } 
\end{eqnarray}

The above relation is completely general, independent of the statistical 
properties of the signal $s \teafi$ and valid for each realization. 
The $\CMtx{A}$ matrix coefficients 
depend on the mask spherical harmonic decomposition and Clebsch-Gordan coefficients.
Then, using the isotropy of the signal, it is possible to compute the coupling matrix
$ \CMtx{ B_{\ell \ell '} }$ relating the $\mathrm{pseudo-}\cl{\ell}$ power spectrum to the true
angular power spectrum.  Although $\CMtx{A} $ is a huge matrix, ($\sim 2 \, 10^6 \times 2 \, 10^6$ elements for 
$\ell_{max} = 1000$), only a small fraction of the elements contribute to 
the $ \CMtx{ B_{\ell \ell '} }$ matrix. 
\begin{equation}
\CVec{ \langle \tilde{ \cl{\ell}^p } \rangle } \eg  \CMtx{ B_{\ell \ell '} } * \CVec{  \cl{\ell'}  } 
\label{Blleq}
\end{equation}

The computations to obtain the $ \CMtx{ B_{\ell \ell '} }$  coupling matrix using the above 
method are rather tedious, while it is possible to compute this matrix simply by applying 
the formalism presented in this paper.

One can construct an estimator $\fcor'$ of the angular correlation function using
equation~(\ref{ksiestfrcl}) with the integrals limited to the partial map where
the function $s(\vec{\Omega})$ is measured.
A lengthy computation \citep{cmvcl} shows that this estimator
is unbiased for separation angles $\gamma < \gamma_{max}$.
However a fast computation of the angular correlation function uses equation~(\ref{ksifrcl})
or its linear form (\ref{ksifrclmat}) and the full sphere function should be used.
This corresponds to equation~(\ref{ksiestfrcl}) with integrals on the full sphere
$S^2 \times S^2$ putting $s(\vec{\Omega})=0$ when $\vec{\Omega}$ is not pointing
to the ``observed'' partial map.
This estimator (noted $\pfcor$), which is of course highly biased,
will be the one used hereafter.

As $\fcor'$ and $\pfcor$ are related by a multiplication by the correlation function
$ \xi^{\mathrm{mask}}(\gamma) $ of the mask itself,
we can notice that the average  of the angular correlation function ($\pfcor$)
computed on the masked sphere is related 
to the correlation function of the complete sphere by:
\begin{eqnarray*}
\langle \fcor'(\gamma) \rangle \ega \xi(\gamma)  \hspace{2mm} \mathrm{(unbiased \,  estimator)} \\
 \pfcor(\gamma)  \ega  \xi^{\mathrm{mask}}(\gamma) \times \fcor'(\gamma) \\
\langle \pfcor(\gamma)  \rangle \ega \xi^{\mathrm{mask}}(\gamma) \times \fcor(\gamma)
\end{eqnarray*}
Here, $\;\xi^{\mathrm{mask}}(\gamma) $ can easily be computed using equation~(\ref{ksifrcl})
and the optimized algorithms cited below, applied to the mask. 
The distortion  matrix in relation (\ref{bctdef}) can then simply be written
as a diagonal matrix $D^{\mathrm{mask}}$
which can be used to compute the $ \CMtx{ B_{\ell \ell '} }$.
\begin{eqnarray}
D^{\mathrm{mask}}(i , i) \ega \xi^{\mathrm{mask}}(\gamma_i) \label{Dmask} \\
\CMtx{ B_{\ell \ell '} }  \ega  \CMtx{K}^{-1} * \CMtx{D^{mask}} * \CMtx{K}  \notag
\end{eqnarray}

\begin{figure}
\includegraphics[width=0.5\textwidth]{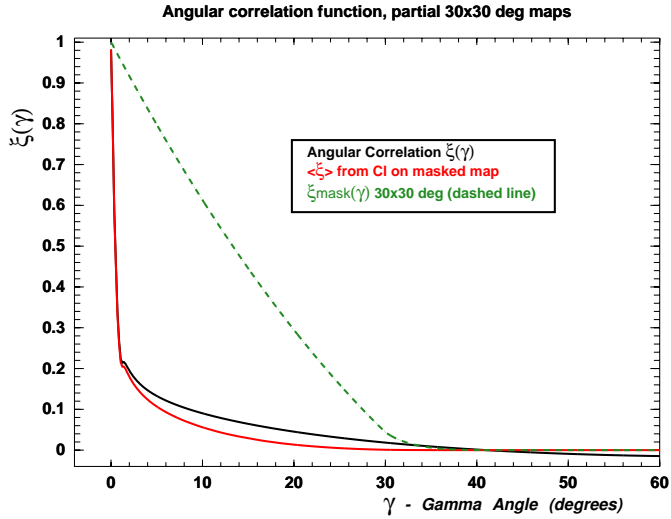} 
\caption{Distortion of the angular correlation function when computed from spherical harmonic decomposition
of a partial (masked) spherical map. The partial map has a $30^\circ$ extension on each of the two orthonal
directions. Rescaled $\xi(\gamma)$ (black) and the distorted $\langle \fcorp \rangle $ (red) are shown for a WMAP-like angular power spectrum, as well the ratio $ \xi^{\mathrm{mask}}(\gamma) =  \langle \fcorp \rangle / \xi^{\mathrm{mask}}(\gamma)$ (dashed green). 
Notice that the angular $\gamma$ range is limited to $60^\circ$ on the figure. }
\label{ximask}
\end{figure}

Figure~\ref{ximask} illustrates the distortion of the angular correlation function estimated from 
$\tilde{a}_{\ell m}$ coefficients computed on the masked map. The undistorted angular correlation 
function correponding to a WMAP-like $\cl{\ell}$ power spectrum, the mean value of the distorted 
correlation function $\langle \fcorp \rangle $ and the mask correlation function $ \xi^{\mathrm{mask}}(\gamma)$ 
are plotted as a function of the $\gamma$ angle, for a partial $30^\circ \times 30^\circ$ map. 

The analysis of the coupling matrix $ \CMtx{ B_{\ell \ell '} }$ shows that the $\ell$-space 
resolution of $\mathrm{pseudo-}\cl{\ell}$ spectrum is compatible with the optimal resolution 
$\Delta \ell \sim \pi / \gamma_{max}$, but has long tails or correlation lengths. These long tails are  
responsible for the systematic shifts of $\mathrm{pseudo-}\cl{\ell}$ spectrum relative to the true one.
This can be seen on figure~\ref{Bllrow} below, showing 
a row of the coupling matrix $B_{\ell \ell '}$ for $\ell=\ell_0=300$.

%%%%%%%%%%%%%%%%%%%%%%%%%%%%%%%%%%%%%%%%%%%%%%%%%%%%%%%
\subsection[]{Partial maps : Correcting $\mathrm{pseudo-}\cl{\ell}$ spectrum }

In some sense, the correlation function  $ \fcorp(\gamma) $ 
obtained through the $\mathrm{pseudo-}\cl{\ell}$ computation applies a weighting 
function ($\xi^{\mathrm{mask}}$) which is inversely proportional to the 
number of available measurement ($s \teafi$) pairs i.e.,
to the statistical significance of the partial sphere correlation function. 
It is possible to obtain the unbiased correlation function by applying the 
inverse correction $1/\xi^{\mathrm{mask}}$ to the correlation function 
of the masked sphere, up to a maximum angle $\gamma_{max}$ which should 
be less than the maximum extent of the partial map. As we increase 
$\gamma_{max}$, the statistical errors on the corrected angular correlation
function, and the corrected angular power spectrum increase.
There is thus a tradeoff between the $\ell$-space resolution and the 
statistical uncertainties of the recovered power spectrum $\clest{\ell}$.  

\begin{figure}
\includegraphics[width=0.5\textwidth]{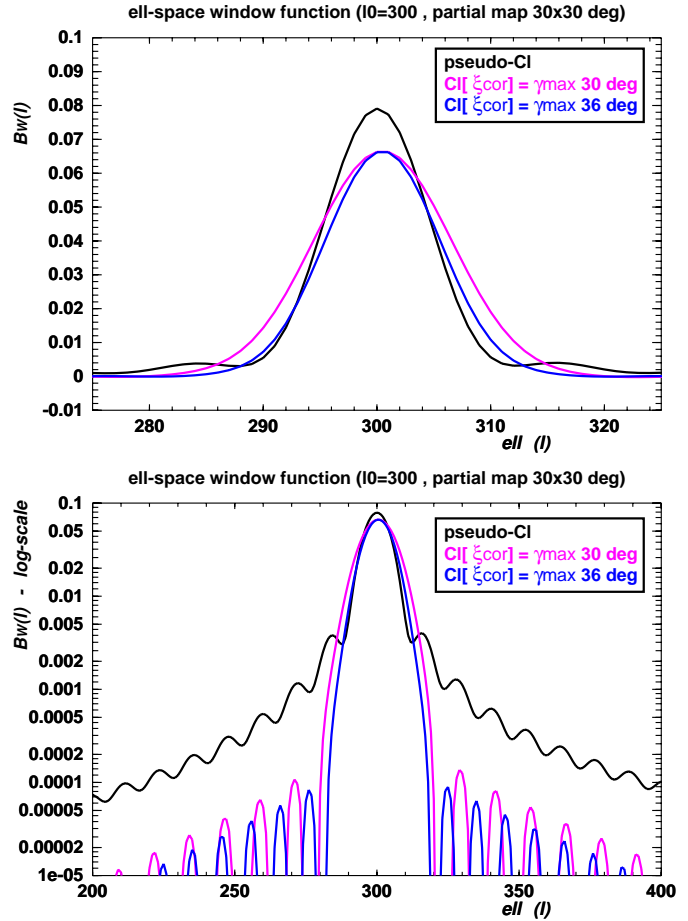} 
 \caption{ $30^\circ \times 30^\circ$ partial map : Comparison of $B_{\ell \ell '}$ matrix row 
for $\ell=300$ (black) with Gaussian weighted $C^t(\ell)$ estimated from 
$\xi^t(\gamma)$, truncated at $\gamma_{max}=30^\circ$ (magenta) and 
$\gamma_{max}=36^\circ$ (Top: linear Y-scale, bottom: logarithmic Y-scale). }
\label{Bllrow}
\end{figure}

The statistical uncertainties (variance) associated with the recovered $\xi$ or $\cl{\ell}$,
depend on the true power spectrum and can not be computed using 
the formalism described here. It is however possible to 
compute the variance for specific partial map shapes, such as polar cap maps (see \cite{cmvcl}).
In this paper, we have  computed 
theses uncertainties using Montecarlo simulation and the corresponding results 
are presented in the next section. 
In figure~\ref{Bllrow}, the $\mathrm{pseudo-}\cl{\ell}$ window function is compared to the 
one obtained for the corrected $\cl{\ell}$ for two values of $\gamma_{max}$
($30^\circ ,\; 36^\circ$). On partial $30^\circ \times 30^\circ$, the maximum possible 
value for $\gamma_{max} \sim \sqrt{2} \times 30^\circ \sim 42^\circ$.
Using a correction up to 
$\gamma_{max}=36^\circ$, we obtain statistical errors comparable 
to those associated with the $\mathrm{pseudo-}\cl{\ell}$, while significantly 
decreasing the systematic effects on the recovered power spectrum
(See section~\ref{Montecarlo} ).

We propose the following procedure to recover an unbiased or corrected 
power spectrum, using fast spherical harmonic decomposition of 
masked spherical maps.
\begin{enumerate}
\item Compute the $\mathrm{pseudo-}\cl{\ell}$ angular power spectrum on the 
masked sphere $\pcl{\ell}$ through fast spherical harmonic decomposition, 
as well as the $\mathrm{pseudo-}\cl{\ell}$ of the the mask itself.
\item Compute the corresponding discrete angular correlation function 
and the mask correlation function 
\begin{equation}
\begin{array}{lcl}
\CVec{ \pfcor(\gamma_i) } \ega \CMtx{K} * \CVec{ \pcl{\ell} }  \\
\CVec{ \fcor^\mathrm{mask}(\gamma_i) } \ega \CMtx{K} * \CVec{ \cl{\ell}^\mathrm {mask} } 

\end{array}
\end{equation} 
\item Define the truncation angle $\gamma_{max}$ compatible 
with the maximum map extent and desired $\ell$ resolution.
Compute the corrected-truncated  angular correlation function using the 
diagonal correction matrix $\mathbf{Dcor}$.

\begin{equation}
\begin{array}{lcl}
 \CVec{ \tfcor(\gamma_i) } \ega \CMtx{Dcor} * \CVec{ \pfcor(\gamma_i) }  \\
 \mathrm{Dcor(i,i)} \ega 1/\xi^{\mathrm{mask}}(\gamma_i)  \hspace{2mm}  \gamma_i \leq \gamma_{max} \\
 \mathrm{Dcor(i,i)} \ega 0 \hspace{2mm} i \ne j \, \mathrm{or} \, \gamma_i > \gamma_{max} 
\end{array}
\end{equation}

The correction matrix is equal to the inverse of the mask distortion (see 
equation~(\ref{Dmask})) if angular correlation information is present 
up to $\gamma_{max}=\pi$. It is useful to apply a  step smoothing function to avoid the 
discontinuity of $\tfcor$ for $\gamma = \gamma_{max}$, such as a sigmoid: 
$$ f(\gamma) = 1/\left[ 1 + \exp \left( \frac{\gamma - \gamma_{max}}{\delta} \right) \right] \hspace{2mm} 
\mathrm{with}  \hspace{2mm}  \delta \sim 0.01 \gamma_{max} $$  
 This step smoothing function  enhances the behavior of the computed power spectrum, 
decreasing residual oscillations. 
\item Compute the corrected angular spectrum and apply the Gaussian
filter function in $\ell$-space with $\sigma_\ell = \pi/\gamma_{max}$.
\begin{equation}
\begin{array}{lcl}
 \CVec{ \tcl{\ell}  } \ega \CMtx{K^{-1}} * \CVec{ \tfcor(\gamma_i) } \\
 \CVec{ \clest{\ell}^t  } \ega  \CMtx{W_G} * \CVec{ \tcl{\ell}  }  \\
 \mathrm{W_G(\ell_0, \ell)} \ega = \frac{ \exp((\ell - \ell_0)^2 / 2 \sigma_\ell^2 ) } { \sum_\ell \exp((\ell - \ell_0)^2 / 2 \sigma_\ell^2 )  } 
\end{array}
\end{equation}
\end{enumerate}

The coupling matrix $\mathrm{B_{c}}$ is independent of the true power spectrum 
and can be computed using the following relation:
\begin{equation}
\begin{array}{lcl}
 \langle \CVec{ \clest{\ell}^t  } \rangle \ega  \CMtx{B_{c}} *  \CVec{\cl{\ell}}  \\
 \CMtx{B_{c}} \ega \CMtx{W_G} * \CMtx{K^{-1}} *  \CMtx{Dcor} *  \CMtx{K} 
\end{array}
\end{equation}

%%%%%%%%%%%%%%%%%%%%%%%%%%%%%%%%%%%%%%%%%%%%%%%%%%%%%%%
%%%%%%%%%%%%                     Section IV                    %%%%%%%%%%%%%
%%%%%%%%%%%%%%%%%%%%%%%%%%%%%%%%%%%%%%%%%%%%%%%%%%%%%%%

\section[]{Simulation results on small maps and masked maps}
\label{Montecarlo}
% As it has been already stated, the formalism presented here can not be 
% used to compute the statistical fluctuations on the computed angular correlation
% function or the power spectrum on incomplete spherical maps.  
There is a tradeoff between the achievable resolution and 
the uncertainties of the estimated power spectrum. In order to evaluate 
these uncertainties, we have performed Montecarlo simulations to generate 
partial and full sky spherical maps and compute the power spectrum
using spherical harmonic decomposition and
the corrected/truncated angular auto correlation function.
{\modif 
The method to correct systematic effects proposed in this work 
is independent of the true power spectrum. The computation of the 
coupling matrix and correction matrix is rather fast with a CPU 
time comparable to few Monte-Carlo realisations using fast spherical
harmonics map analysis ($\sim$ minute). Reasonable estimates of errors
for a given input power spectrum can be obtained using few hundred
realisations. 
Although it is possible to compute the systematic effects due to 
the limited sky coverage using Monte-Carlo, this would require much larger computation
time compared to the uncertainty estimates. The estimate of the  
$\CMtx{ B_{\ell \ell '} } $ coupling matrix elements with sufficient precision 
would need a large number of Monte-Carlo realisations, due to the 
cosmic variance. In addition, this computation process would have
to be repeated for several  input power spectra.
} 

\begin{figure}
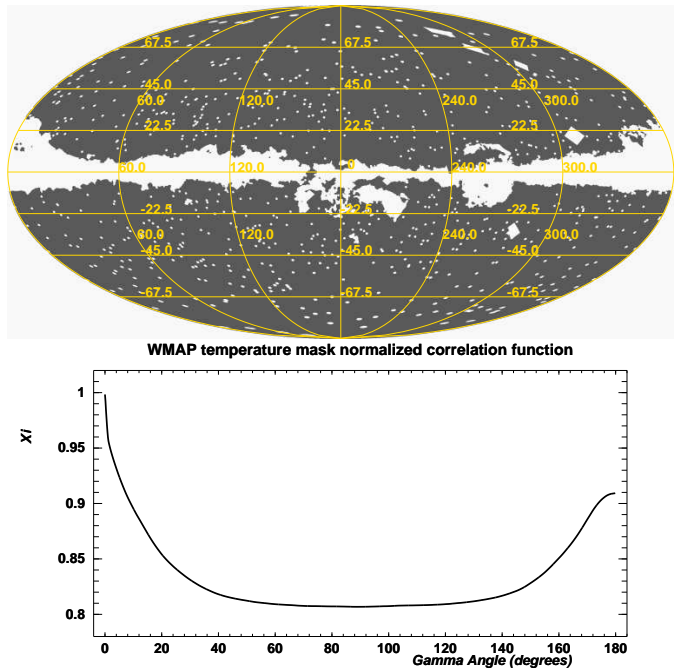

\includegraphics[width=0.5\textwidth]{\grfile{maskwmap}} \\
\includegraphics[width=0.5\textwidth]{\grfile{ximaskwmap}} 
 \caption{ WMAP KQ85 CMB temperature analysis mask (top) and the corresponding 
normalised correlation function $\xi^{\mathrm{mask}}(\gamma)$ (bottom).  }
\label{maskwmap}
\end{figure}

The simulations have been performed with different partial map 
geometries and input (true) $\cl{\ell}$ power spectra to check the validity of the 
conclusions. However, for the sake of clarity,
we present here the results only for two map geometries and two input spectra.
We have used WMAP-like true $\cl{\ell}$ power spectrum, 
labeled $t_{wmap}$  and a second shape, labeled  $t_6$, which have sharp features 
on top of a smooth  spectrum $C(\ell) = Cte/[\ell\times(\ell+1)] + \mathrm{peaks}$ in order to illustrate the 
$\ell$-space resolution effect. The results presented here have been obtained
for the following two map geometries:
\begin{enumerate}
\item Partial square map, $30^\circ\x 30^\circ$, covering the angular 
range $75^\circ \leq \theta \leq 105^\circ$ and $ \phi_0 \leq \phi \leq \phi_0 + 30^\circ$.
This map covers $\sim 0.27\;sr\;=\; 0.022\;\times \;4 \pi\;sr$.
This small map corresponds to the case of ground or balloon CMB instruments.

\item A nearly complete spherical map, but with an equatorial band removed (set to zero).
This case shows the distortion of the  $C(\ell)$ power spectrum obtained from all sky 
space experiments such as WMAP or Planck, where some part of the sky with strong non-CMB 
microwave signals has to be masked. The removed equatorial band illustrates the effect
of Galactic cut applied to CMB maps. We have performed simulations on $4 \pi$ spherical maps
with a $\Delta \theta = \pm 10^\circ$ equatorial band cut 
($s \teafi = 0$  for $80^\circ \, < \theta < 100^\circ\; , \forall \phi$).
Such a map covers 82.5 \% of the whole sky ($0.825 \times 4 \pi\; sr$) and provides 
correlation information for the whole angular range ($ 0 \leq \gamma \leq \pi $).  

\item { \modif A more realistic case where we have applied the 5-year WMAP CMB temperature
analysis (KQ85) mask \citep{galwmap} to simulated CMB maps. The mask and corresponding normalised 
correlation function is shown in figure~\ref{maskwmap}.
The unmasked area corresponds to $\sim 0.82 \times 4 \pi\; sr$, nearly equal to 
the area of maps with a $\pm 10^\circ$ equatorial band removed. It has a complex 
and patchy shape.  }
\end{enumerate} 

Given the overall shape of the $\cl{\ell}$ spectra studied here, 
we have represented $\ell \times (\ell+1) \, \cl{\ell}$ for the power spectrum, and 
$\ell \times (\ell+1) \, \sigma_{\cl{\ell}}$ for the associated 
statistical uncertainties on all figures that follow in this section.
{\modif 
The systematic shifts of the recovered power spectrum can be quantified  
using difference between the recovered and true power spectrum, normalised to 
the cosmic variance $ \sigma_{cv}(\ell) $ 
$$ \delta ( \ell ) =  \frac{\langle \clest{\ell} \rangle - \cl{\ell} }{  \sigma_{cv}(\ell) }  $$
} 
\subsection[]{Small $30^\circ\times 30^\circ$ maps} 
\begin{figure}
\includegraphics[width=0.5\textwidth]{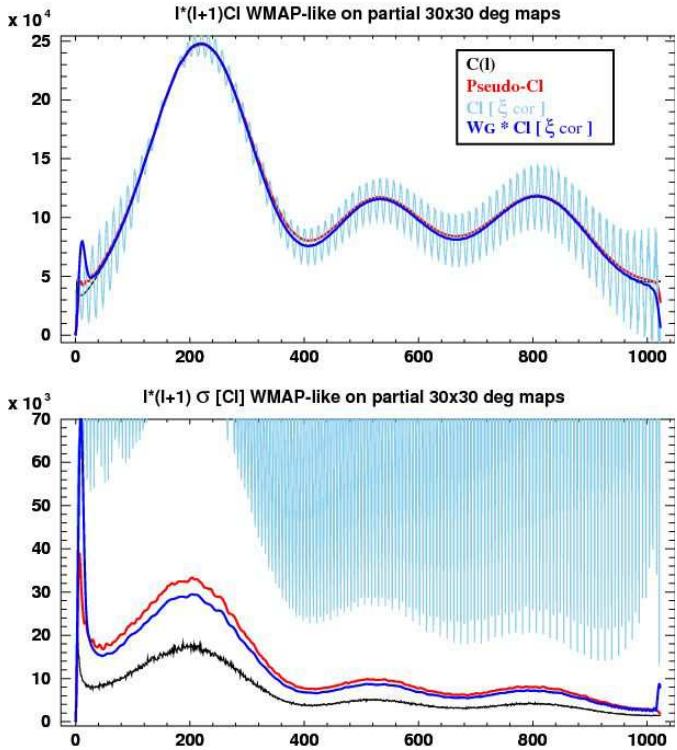} 
 \caption{ $30^\circ \times 30^\circ$ partial maps : Comparison of computed power spectrum
and true $t_{wmap}$ power spectrum (top) and the associated statistical errors (bottom).
True power spectrum in black, $\mathrm{pseudo-}\cl{\ell}$ in red, $\cl{\ell}^t$ computed 
from corrected angular correlation function truncated at $\gamma_{max} = 30^\circ$ (light blue), and corrected 
$C(\ell)^t$ binned with Gaussian weights (blue)  }
\label{clms1wmap30x30}
\end{figure}

\begin{figure}
\includegraphics[width=0.5\textwidth]{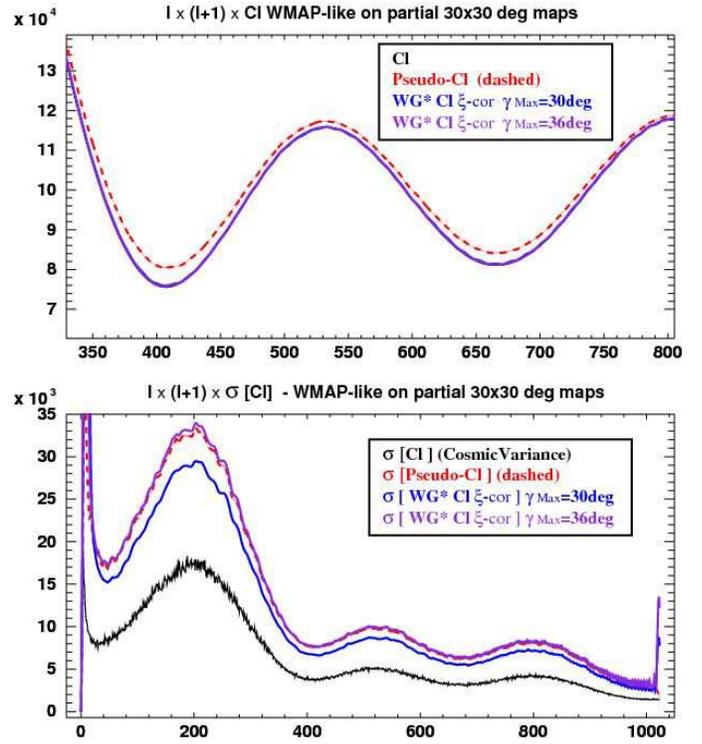} 
 \caption{ $30^\circ \times 30^\circ$ partial maps : Comparison of computed power spectrum
and true $t_{wmap}$ power spectrum (top)  and the associated statistical errors (bottom).
True power spectrum in black, $\mathrm{pseudo-}\cl{\ell}$ in red, 
Gaussian weighted binned $\cl{\ell}^t$ computed from corrected angular 
correlation function truncated at $\gamma_{max} = 30^\circ$ (blue), and with  $\gamma_{max} = 36^\circ$ (violet)  }
\label{clms2wmap30x30}
\end{figure}

The figures~\ref{clms1wmap30x30} and~\ref{clms2wmap30x30} show the 
recovered power spectrum on small $30^\circ\times30^\circ$ maps for a true WMAP-like
power spectrum. The systematic shifts of the $\mathrm{pseudo-}\cl{\ell}$ power spectrum 
are clearly visible, in particular on figure~\ref{clms2wmap30x30} (top). 
The false oscillations on the raw power spectrum recovered from truncated 
$\xi$ is also shown on the figure~\ref{clms1wmap30x30} (top).

\begin{figure}
\includegraphics[width=0.5\textwidth]{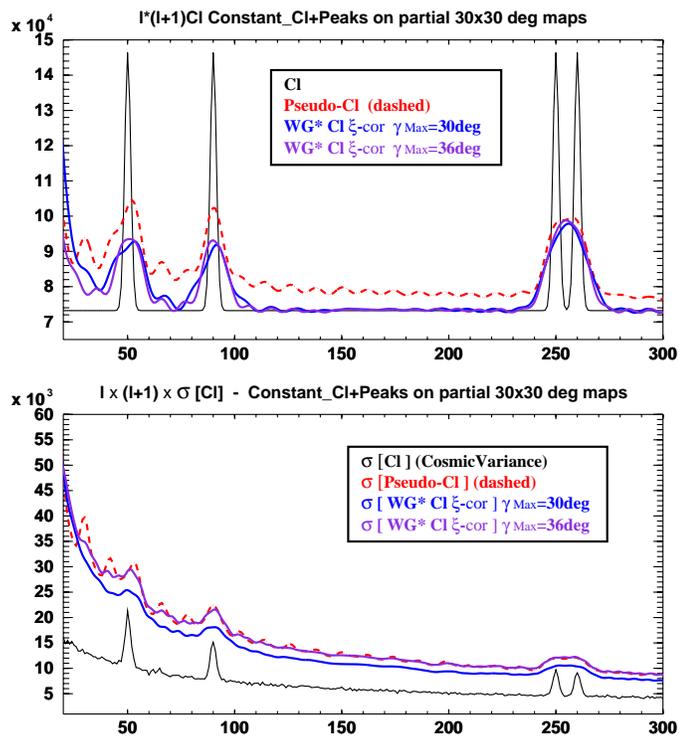} 
 \caption{ $30^\circ \times 30^\circ$ partial maps : Comparison of computed power spectrum 
and true $t_6$ power spectrum (top)  and the associated statistical errors (bottom).
True power spectrum in black, $\mathrm{pseudo-}\cl{\ell}$ in red, 
Gaussian weighted binned $\cl{\ell}^t$ computed from corrected angular 
correlation function truncated at $\gamma_{max} = 30^\circ$ (blue), and with  $\gamma_{max} = 36^\circ$ (violet)  }
\label{clms2t630x30}
\end{figure}

It can also be seen that by choosing $\gamma_{max} = 36^\circ$, it is possible 
to avoid nearly all the systematic shifts of the power spectrum, while 
keeping the level of statistical fluctuations comparable to the $\mathrm{pseudo-}\cl{\ell}$,
and without losing the $\ell$-space resolution (See figure~\ref{clms2wmap30x30} 
and~\ref{clms2t630x30}). However, it should also be noted that the 
corrected $\cl{\ell}$ suffers a systematic overestimation at low 
$\ell \lesssim \left( 2-3 \times \sigma_\ell= 2-3 \pi/\gamma_{max} \right)$, as expected from the coupling matrix $\mathrm{B_{c}}$.

{ \modif
The systematic shift $\delta(\ell)$ decreases by a factor 10-30, changing 
from $\delta \sim 0.25-0.75$ in the case of $\mathrm{pseudo-}\cl{\ell}$ to 
less than 0.05 for Gaussian weighted $\cl{\ell}^t$ computed from $\xi(\gamma)$ 
corrected up to $\gamma_{max} = 36^\circ$ (figure~\ref{clms2wmap30x30}).
}

\subsection[]{Maps with $\pm$ $10^\circ$ Galactic cut }  
\label{pm10gc}
\begin{figure}  
\includegraphics[width=0.5\textwidth]{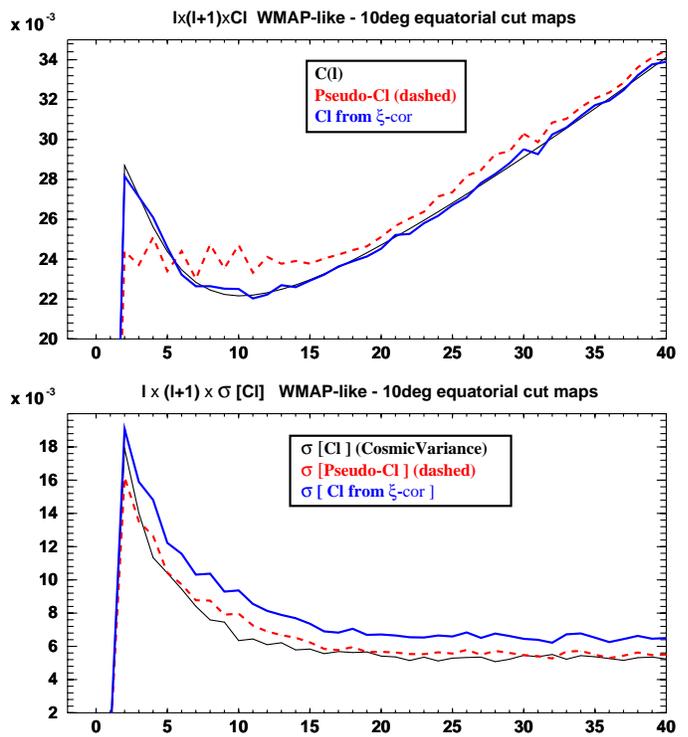} 
\caption{ $4 \pi$ maps with $\pm\; 10^\circ$ equatorial cut: Comparison of computed power spectrum 
and true $t_{wmap}$ power spectrum (top)  and the associated statistical errors (bottom).  
True power spectrum in black, $\mathrm{pseudo-}\cl{\ell}$ in red, 
Gaussian weighted binned $\cl{\ell}^{\xi c}$ computed from corrected angular correlation function in blue.  }
\label{clmswmapgc10}
\end{figure}

\begin{figure}
\includegraphics[width=0.5\textwidth]{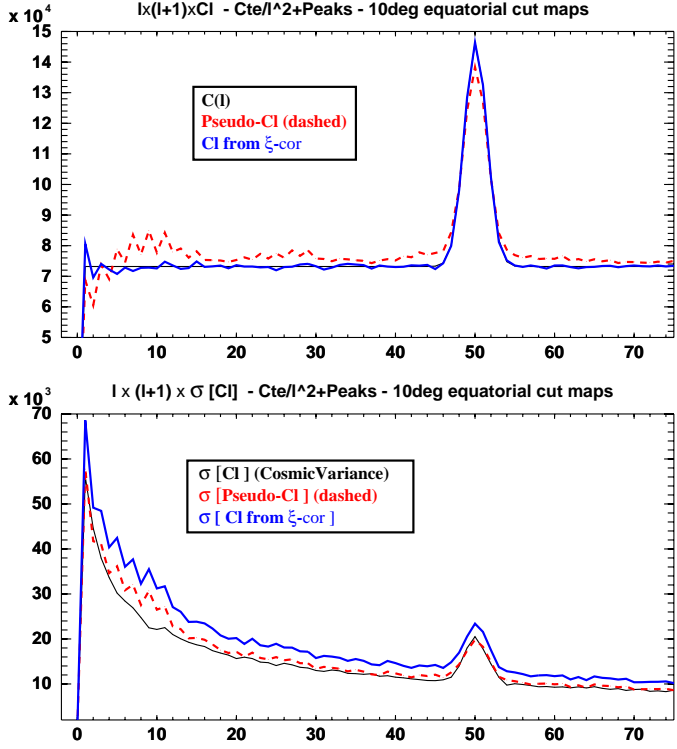} 
\caption{ $4 \pi$ maps with $\pm\;10^\circ$ equatorial cut: Comparison of computed power spectrum
and true $t_6\;$ ($C(\ell) = \mathrm{Cte}/\ell\times(\ell+1) + \mathrm{peaks}$) power spectrum (top)  and the 
associated statistical errors (bottom). True power spectrum in black, $\mathrm{pseudo-}\cl{\ell}$ in red, 
$\cl{\ell}^{\xi c}$ computed from corrected angular correlation function in blue  }
\label{clmst6gc10}
\end{figure}

The figure~\ref{clmswmapgc10} shows the bias introduced on the recovered power spectrum,
specially at low multipoles $\ell \lesssim 30$ using uncorrected $\clest{\ell}^p$ from 
map projection on the $\ylm{\ell}{m}$ basis. In this case, the angular correlation function 
can be estimated at all angular scales, but the $\fcorp(\gamma)$ obtained from the 
$\mathrm{pseudo-}\cl{\ell}$ is distorted.
By correcting the  $\fcorp(\gamma)$, it is possible to recover 
the unbiased power spectrum $\clest{\ell}^{\hat{\xi}}$,
as shown on figure~\ref{clmswmapgc10}  (top).
{ \modif
The systematic shift $\delta(\ell)$ changes from $\delta \sim 0.25$ in the 
case of $\mathrm{pseudo-}\cl{\ell}$ to less than 0.05 for Gaussian weighted $\clest{\ell}$ 
computed from corrected $\xi(\gamma)$.  
}
However, the statistical fluctuations are larger, compared to the cosmic variance,
reachable if the complete $4 \pi$ map is available, or to the $\mathrm{pseudo-}\cl{\ell}$ variance 
on the cut map. Figure~\ref{clmst6gc10} shows the effect of the equatorial 
cut on a sharp feature present in the simulated power spectrum around $\ell =50$.
The power spectrum computed from corrected angular autocorrelation also corrects
this distortion. 

Although more sophisticated methods can be used to recover the spectrum at low 
multipoles from cut maps, the correction 
algorithm proposed here can be used to easily correct the $\mathrm{pseudo-}\cl{\ell}$ spectrum.

\subsection[]{ \modif Maps with WMAP KQ85 mask  }   
\begin{figure}  
\includegraphics[width=0.5\textwidth]{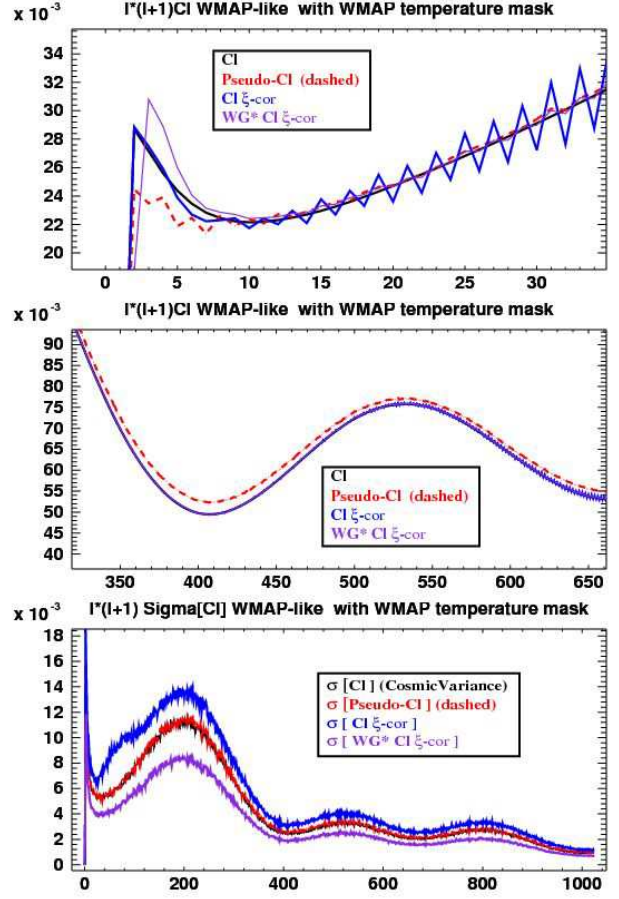} 
\caption{ $4 \pi$ maps masked with the WMAP 5 year temperature KQ85 mask. 
Comparison of recovered power spectrum and true $t_{wmap}$ power spectrum (top and middle)
and the associated statistical errors (bottom).  
True power spectrum in black, $\mathrm{pseudo-}\cl{\ell}$ in red, dashed,  
$\cl{\ell}^{\xi c}$  computed from corrected angular correlation function in blue, and 
filtered with Gaussian weights ($\sigma_\ell = 0.75$) in violet  $\cl{\ell}^{\xi c,G}$.  }
\label{clmswmapmask}
\end{figure}

{ \modif 
The effect of the WMAP KQ85 foreground suppression mask on the recovered power spectrum 
is shown in figure~\ref{clmswmapmask}. This mask cuts an equatorial area on the map
which is significantly less extended than in the case of the simple $\pm$ $10^\circ$ cut discussed 
above, but it removes also a large number of smaller patches of the sky scattered  over the
map.  As a result, it can be seen that the systematic effects at low multipoles on the 
recovered $\mathrm{pseudo-}\cl{\ell}$ are smaller compared to the more extended 
$\pm\;10^\circ$ equatorial cut, while significant shift appears at higher multipoles,  
around $\ell = 400$ for instance, due to the patchy structure of the mask.
This patchy structure is also responsible for the oscillatory behaviour of the 
$\cl{\ell}^{\xi c}$ power spectrum computed from the corrected 
angular correlation function, for $\ell \gtrsim 10$. Theses unphysical 
oscillations can be smoothed out by applying a narrow Gaussian ($\sigma_\ell = 0.75$)
filter function, $\cl{\ell}^{\xi c,G}$,  as can be seen on figure~\ref{clmswmapmask}.

As expected, correcting $\xi(\gamma)$ using $\xi^{\mathrm{mask}}(\gamma)$ increases 
the statistical uncertainties compared to $\mathrm{pseudo-}\cl{\ell}$, while the filtered
power spectrum has smaller variance. It is possible to use the combined power spectrum,
$\cl{\ell}^{\xi c}$  without filtering for $\ell \lesssim 10$ and Gaussian weighted $\cl{\ell}^{\xi c,G}$
for $\ell \gtrsim 10$.
The systematic shift $\delta(\ell)$ decreases  from $\delta \sim 0.5-1.5$ in the 
case of $\mathrm{pseudo-}\cl{\ell}$ to $\delta \lesssim 0.05$  for this combined $\clest{\ell}$. 

We have also checked that our conclusions are valid in the presence of noise.
We have performed simulations where Gaussian fluctuations, with smooth spatial variations
have been added to simulated maps.
The reconstructed power spectrum corresponds to the sum of the sky signal 
and the noise spectra ($ \clest{\ell} = \cl{\ell} + \cl{\ell}^{\mathrm{noise}} $), distorted by the 
mask, as it is expected for nearly isotropic uncorrelated noise.
}

\section[]{Conclusions}
We have established a simple method to evaluate and correct systematic effects associated 
with power spectrum computed using the $\mathrm{pseudo-}\cl{\ell}$ on partial maps. 
The coupling matrix $B_{\ell \ell'}$ can indeed be computed using matrix algebra and the 
sky mask spherical harmonic decomposition. We have also described an 
algorithm to correct the systematic shifts of the calculated power spectrum, as well 
as a near optimal $\ell$-space window or filter function. 
It should be noted that it is possible to improve the $\ell$ resolution,
compared to the {\it natural} resolution $\Delta \ell \sim \sigma_\ell \sim \pi / \gamma_{max}$
at the expense of higher statistical fluctuations.
For  all sky CMB experiments such as WMAP or Planck,
non-CMB dominated parts of the sky (galaxy \ldots)  are usually excluded 
from the angular power spectrum estimation. We show that our method can be used to correct 
the power spectrum distortions at low $\ell$ in such cases, when the angular power 
is computed using fast spherical  harmonic decomposition on almost complete maps.

The correction method described here can easily be extended to the CMB-polarisation
power spectrum. A similar approach for polarised maps has been developed in \citep{ksclpol}.
It should also be possible to use this method to improve the angular power 
spectrum calculation by taking into account individual pixel measurement errors.  
For observations with negligible correlated noise, a weighting function mask inversely 
proportional to the individual pixel measurement uncertainties can be applied to the map 
before decomposition on the $Y_{\ell m}$ basis, and subsequently corrected for.

\section*{Appendix}
\label{appmatinv}

We will show that one can invert equation~(\ref{ksifrclmat}) for a certain choice of
the separation angles $\gamma$.
We assume that the power at large $\ell$ is negligible.
We can then consider that $\cl{\ell }=0$ for all multipoles with $\ell > \ell_{max}$.
The sum in equation~(\ref{ksifrcl})
becomes finite, so 
for $n_\gamma$ given discrete values of $\gamma$,
the matrix elements in equation~(\ref{ksifrclmat}) are:
\begin{equation}
[[K]]_{i\ell } \eg \frac{2\ell+1}{4\pi}\;\pl{\ell}(\cos(\gamma_i))
\end{equation}
with \; $ 1\le i\le n_\gamma $ \; and \; $0\le \ell\le \ell_{max}$.

If the matrix $[[K]]$ is invertible, one may write
$\;[\cl{\ell }] = [[K]]^{-1}*[\fcor_i]\;$ and the $\cl{\ell }$
spectrum can be recovered from the values of the
angular correlation function computed at angular separations
$\gamma_i$. \\
In the following we will show that the separation angles $\gamma_i$
can be choosen such that the square matrix is non-singular.

$\pl{\ell }(x=\cos(\gamma))$ is a polynomial of degree $\ell$
in the interval $[-1,+1]$.
Using equation~(\ref{ksifrcl}) we see that $\fcor(\gamma)$
is a polynomial of degree at most $\ell_{max}$.
The integrand $\fcor\x \pl{\ell }$ of equation~(\ref{clfrksi})
is a polynomial of degree at most $\ell_{max}+l$,
so for the highest multipole
$\cl{\ell_{max}}$ the integrand has a degree at most $2 \ell_{max}$.

We know that, using Gauss-Legendre quadrature of order $n$,
integrals of polynomials of degree up to $2n+1$ can be
exactly expressed as a $n$-term weighted sum of the polynomials
computed at special values of $x$ (see ~\cite{abra72}).

For all $\ell\le \ell_{max}$ and for $L \ge \ell_{max}+1$, the integral in 
equation~(\ref{clfrksi}) can be expressed as the algebraic sum:
\begin{equation}
\cl{\ell } \eg
 2\pi \; \somme{i=1}{L}\;\fcor(\gamma_i)\pl{\ell }(\cos(\gamma_i))\x \;w_i
\end{equation}
where $x_i=\cos(\gamma_i)$ are the $L$ roots of $\pl{L }(x)$
and $w_i$ the weights. For $L \geq \ell_{max}+1$, 
this can be written in a matrix form:
\[ \begin{array}{rclcl}
[\cl{\ell }] &=& [[K']]*[\fcor_i]
        &\qquad & 0\le \ell\le \ell_{max} \\
&&&& \\
K'_{\ell i} &=& 2\pi\pl{\ell }(\cos(\gamma_i))\x \;w_i
      &\qquad & 1\le i\le L 
\end{array} \]
Clearly, if we take $L=\ell_{max}+1$,
the number of abscissas values is $n_\gamma=\ell_{max}+1$
and the matrix $[[K]]$ is square.
In that case $[[K']]=[[K]]^{-1}$ and equation~(\ref{clfrksi}) can be written
\begin{equation}
[\cl{\ell }] = [[K]]^{-1} *[\fcor_i]
\quad \left( 1 \le i \le \ell_{max}+1  , \; 0\le \ell \le \ell_{max} \right)
\end{equation}
\noindent  { \bf Remark:} 
Numerically, the angular correlation function could also be computed
at $\ell_{max}+1$ regularly spaced values $\gamma_i$ between $0$ and $\pi$.
This is possible because the $\gamma_i$ corresponding to
the Legendre polynomial roots $x_i=\cos(\gamma_i)$ satisfy the relation 
\begin{equation}
\frac{2i-1}{2\ell+1} \;\le\; \frac{\gamma_i}{\pi} \;\le\; \frac{2i+1}{2\ell+1}
\qquad\text{for}\quad\pl{\ell }(\cos(\gamma))
\end{equation}
and are nearly regularly spaced.

Thus the angular distance between a Legendre polynomial root and the
nearest regularly spaced abscissa is always lower than the
resolution associated with the highest multipole ($\sim \pi / \ell_{max} $).
\section*{Acknowledgments}

We thank  J. Haissinski, J.P. Pansart and J. Rich for useful discussions and comments.

\bsp

\label{lastpage}

\end{document}